\documentclass[prc,preprint,showpacs,tightenlines,%
nofootinbib]{revtex4}

\usepackage{graphicx}  
\usepackage{bm}  
\usepackage{amsmath}
\usepackage{epsf}
\newcommand{\beq}{\begin{equation}}
\newcommand{\eeq}{\end{equation}}
\newcommand{\bea}{\begin{eqnarray}}
\newcommand{\eea}{\end{eqnarray}}

\newcommand{\simgt}{\stackrel{>}{{}_\sim}}

\begin{document}
\preprint{HISKP-TH-06/31}
\title{Universal Properties of the Four-Body System 
with Large Scattering Length}
\author{H.-W. Hammer}\email{hammer@itkp.uni-bonn.de}
\affiliation{Helmholtz-Institut f\"ur Strahlen- und Kernphysik (Theorie),
Universit\"at Bonn, 53115 Bonn, Germany}
\author{L. Platter}\email{lplatter@phy.ohiou.edu}
\affiliation{Department of Physics and Astronomy,
Ohio University, Athens, OH 45701, USA\\}
\date{\today}
\begin{abstract}
Few-body systems with large scattering length have universal 
properties that do not depend on the details of their interactions
at short distances. We study the universal bound state properties of the 
four-boson system with large scattering length in an effective 
quantum mechanics approach. We compute the four-body binding energies 
using the Yakubovsky equations for positive and negative scattering length.
Moreover, we study the correlation between three- and four-body
energies and present a generalized Efimov plot for the  four-body
system. These results are useful for understanding the cluster structure 
of nuclei and for the creation of weakly-bound tetramers with cold atoms
close to a Feshbach resonance.
\end{abstract}
\pacs{03.65.Ge, 36.40.-c, 21.45.+v}
\keywords{}
\maketitle
\section{Introduction}
\label{sec:intro}
Effective theories are a powerful framework to exploit a separation
of scales in a physical system.
The long-distance degrees of freedom must be included dynamically 
in the effective theory, while short-distance physics enters only 
through the values of a few coupling constants, often called
low-energy constants.
Effective theories are widely used in many areas of physics.  
Recently, a considerable effort was devoted to applying
effective field theories in nuclear and atomic physics
\cite{Bedaque:2002mn,Epelbaum:2005pn,Braaten:2004rn}.
If there is no exchange of massless particles, any interaction will
appear short ranged at sufficiently low energy. One can then use
a very general effective theory with short-range interactions only
to describe the universal low-energy properties of the system. 
Such a theory can be applied to a wide range of systems 
from nuclear and particle physics to atomic and molecular physics.

In this paper, we focus on
few-boson systems with a large two-body scattering length.
They are characterized by an unnaturally large two-body scattering length $a$
which is much larger than the typical low-energy length scale $l$ given
by the range of the interaction. Such systems
display a number of interesting effects and universal properties that are 
independent of the details of the interaction at short distances of order $l$. 
The simplest one is the existence of a shallow two-body bound state 
(called dimer) with universal binding energy 
$B_2=\hbar^2/(ma^2)+{\cal O}(l/a)$
for positive $a$ and where $m$ is the mass of the particles. 
Low-energy observables can generally be described in a controlled expansion 
in $l/|a|$. In the two-boson system, the effective 
theory reproduces the effective range expansion but
the structure of the three-boson system is much richer \cite{Braaten:2004rn}.
It has universal properties that include a geometric spectrum of 
three-body bound states (so-called Efimov trimers),
log-periodic dependence of three-body observables on the scattering length,
and a discrete scaling symmetry \cite{Efimov71,Efimov79}. 
These features can be understood as the consequence of a  
renormalization group limit cycle in the three-body system
\cite{Bedaque:1998kg,Bedaque:1998km,Mohr:2005pv}. 

In the effective field theory (EFT) formulation of Bedaque et 
al.~\cite{Bedaque:1998kg,Bedaque:1998km}, 
the limit cycle is manifest in the renormalization group behavior
of a contact three-body interaction required for consistent renormalization.
This implies that at leading order in $l/|a|$, the properties of the 
three-boson system with large scattering length are not determined by 
two-body data alone and one piece of three-body information 
(such as a trimer energy) is required as well.
This three-body information can be specified in various ways
\cite{Braaten:2004rn,Bedaque:1998kg,Bedaque:1998km,Mohr:2005pv,Afnan:2003bs,Platter:2006ev,Platter:2006ad,Platter04}. 
In the following, we will use the parameter $L_3$ introduced
in \cite{Platter04}.
As an alternative to the EFT formulation, one can construct an 
effective theory in a quantum mechanical framework \cite{Lepage:1997cs}.
The contact operators in the field theory are then replaced by 
an ``effective potential'' built from smeared out
$\delta$-function potentials and derivatives thereof. 
The qualitative features of the renormalization are very similar to the 
EFT formulation and also show the limit cycle behavior.
In the case of positive scattering length, this approach has 
successfully been applied to the three-
and four-boson systems ~\cite{Mohr:2003du,Mohr:2005pv,Platter04}
and to the four-nucleon system \cite{Platter:2004zs}.
Here we extend this work to negative scattering lengths.

The four-body problem has previously been studied in more traditional
approaches. Early studies include the Schr\"odinger equation 
with separable potentials \cite{Gibs76} and field-theoretical models with
separable expansions of the three-body T-matrix \cite{Fons76}.
For a review of these and other early studies, see 
Refs.~\cite{Tjon78,Lim84}. The nuclear four-body problem has recently
been benchmarked by comparing various modern calculational approaches
\cite{Kama01}. An overview of recent calculations for the four-body system
of $^4$He atoms can be found in Ref.~\cite{Blume:2000}.

None of these previous works focused on the universal properties of the 
four-body system with large scattering length which remain poorly
understood. A first step towards understanding the four-body system with
large positive $a$ was taken in Ref.~\cite{Platter04}.
By means of an explicit calculation, it was demonstrated
that the renormalization of the three-body system automatically 
ensures the renormalization of the four-body system
in this case. Therefore no four-body parameter should enter at leading order. 
These results were applied to calculate the $^4$He tetramer 
ground and excited state energies and good agreement with the 
Monte Carlo results of Blume and Greene was found \cite{Blume:2000}.  
Within the renormalized zero-range model, however,
Yamashita et al.~\cite{Yama06} recently found a strong sensitivity
of the deepest four-body energy
to a four-body subtraction constant in their equations.
They motivated this observation from a general model-space reduction
of a realistic two-body interaction close to a Feshbach resonance.
The results of Ref.~\cite{Platter04} for the $^4$He tetramer
were also reproduced. Yamashita et al. concluded that a four-body 
parameter should generally enter at leading order. They argued that 
four-body systems of $^4$He atoms and nucleons (where this sensitivity
is absent \cite{Platter04,Platter:2004zs,Nogga:2004ab}) 
are special because repulsive interactions strongly reduce 
the probability to have four particles close together.
However, the renormalization of the four-body problem was not explicitly 
verified in their calculation.
Another drawback of their analysis is the focus on the deepest four-body
state only. Therefore, it remains to be seen 
whether their findings are universal or are an artefact of 
their particular regularization scheme.

The purpose of this paper is to extend the study of 
Ref.~\cite{Platter04} in two important ways. First, we carry out
calculations for positive and negative scattering length and 
study the universal correlations between the three- and four-body
binding energies. Second, we map out the scattering 
length dependence of the shallowest two four-body states and 
summarize the spectrum in a generalized Efimov-plot. 
We will follow Ref.~\cite{Platter04} and
work at leading order in $l/|a|$ using the framework of 
non-relativistic quantum mechanics to construct an effective interaction 
potential. The universal properties of the four-body spectrum
will be useful for understanding the cluster structure 
of nuclei \cite{Mar06} as part of the planned experimental program at FAIR
and for the creation of weakly-bound tetramers with cold atoms
close to a Feshbach resonance. A first step in this direction was 
already taken in  Ref.~\cite{Chin05}.

The organization of the paper is as follows: In Sec.~\ref{sec:effT},
we briefly describe our formalism for the two-, three-, and four-body
bound state equations. 
The discussion of the universal correlations and the four-body
spectrum follows in Sec.~\ref{sec:res}. Finally, we summarize and
present an outlook in Sec.~\ref{sec:conc}.

\section{Few-Body Bound State Equations in Effective Theory}
\label{sec:effT}
The effective low-energy interaction potential generated by a 
non-relativistic EFT with short-range interactions can be written
down in a momentum expansion. In the two-body S-wave sector,
it takes the general form
\beq
\langle {\bf k'} | V | {\bf k} \rangle =
\lambda_2 + \lambda_{2,2} (k^2+k'^2)/2 +\ldots\,,
\label{effpot}
\eeq
where ${\bf k}$ and ${\bf k'}$ are the relative three-momenta of
the incoming and outgoing particles, respectively.
Similar expressions can be derived for three- and higher-body 
interactions. The exact form of the potential depends on the specific
regularization scheme used. The low-energy observables, however, 
are independent of the regularization scheme (up to higher order corrections)
and one can choose a convenient scheme for practical calculations.

We regularize the potential in Eq.~(\ref{effpot}) by multiplying with 
a Gaussian regulator function, $\exp[-(k^2+k'^2)/\Lambda^2]$, with
the cutoff parameter $\Lambda$. This factor strongly suppresses 
high-momentum modes in the region $k,k' \simgt \Lambda$ where the 
effective potential is not valid. 
The cutoff dependence of the coefficients $\lambda_2(\Lambda)$, 
$\lambda_{2,2}(\Lambda)$, $\ldots$ is determined by the requirement
that low-energy observables are independent of $\Lambda$. 

The expansion in Eq.~(\ref{effpot}) is only useful in conjunction with a 
power counting scheme that determines the relative importance of the various
terms at low energy.
The leading order is given by the $\lambda_2$ term which must 
be iterated to all orders, while the other terms give rise to higher-order
corrections that can be included perturbatively  
\cite{Kaplan:1998tg, vanKolck:1998bw}.
In this paper, we will restrict ourselves to leading order in the 
expansion in $l/|a|$ and include only the $\lambda_2$ term.
In the three-body system, 
a momentum-independent three-body interaction term $\lambda_3$ must be
included together with $\lambda_2$ at leading order  in $l/|a|$
\cite{Bedaque:1998kg,Bedaque:1998km}. 
Effective range effects and other higher-order corrections can be 
included as well 
\cite{Hammer:2001gh,Bedaque:2002yg,Afnan:2003bs,Griesshammer:2004pe,Platter:2006ev,Platter:2006ad}.
In the four-body system, no new parameter enters at leading order
in $l/|a|$ and only $\lambda_2$ and $\lambda_3$ contribute
\cite{Platter04}. 

In order to set up our conventions and formalism,
we will briefly review the bound state equations for the two-,
three-, and four-body systems. For a more detailed discussion including
explicit equations in momentum space, we refer the reader to
Refs.~\cite{Platter04,Lucthesis}.

\subsection{The Two-Body Sector}
\label{sec:effT:2body}
We write the leading order two-body effective potential in momentum space as:
\begin{equation}
\langle {\bf p}|V|{\bf q}\rangle=
\langle {\bf p}|g \rangle \lambda_2 \langle g|{\bf q} \rangle ~,
\label{effpot_2}
\end{equation}
where $\lambda_2$ denotes the two-body coupling constant and
${\bf q}$ (${\bf p}$) are the relative three-momenta in the incoming
(outgoing) channel. The regulator functions
\beq
\langle {\bf p}|g\rangle \equiv g(p)=\exp(-p^2/\Lambda^2)~,
\eeq
suppress the contribution from high momentum states. In the few-body 
literature, they are often called ``form factors''. 
For convenience, we will work in units where
Planck's constant $\hbar$ is set to unity: $\hbar=1$.

The interaction (\ref{effpot_2}) is separable and the Lippmann-Schwinger
equation for the two-body problem can be solved analytically.
The two-body t-matrix can be written as \cite{Ziegelmann}:
\begin{equation}
\label{t-matrix}
t(E)=|g\rangle\tau(E)\langle g|~,
\end{equation}
where $E$ denotes the total energy. The two-body propagator $\tau(E)$ is
then given by
\begin{equation}
\tau(E)=\left[1/\lambda_2-4\pi\int_0^\infty\hbox{d}q\,
  q^2\frac{g(q)^2}{mE-q^2}\right]^{-1}~.
\end{equation}

A two-body bound-state appears as a simple pole in the two-body propagator
$\tau$ at energy $E=-B_2$. Thus for $a>0$, the two-body coupling constant 
$\lambda_2(B_2,\Lambda)$ can be fixed from the binding energy $B_2$,
which is directly related to the scattering length by $a=1/\sqrt{mB_2}$
up to higher order corrections in $l/|a|$.
For negative scattering length, there is no shallow dimer
state and the  coupling constant $\lambda_2$ is fixed from matching to
the effective range expansion for positive energies.

%
\subsection{The Three-Body Sector}
\label{sec:effT:3body}
The low-energy properties of the three-body system for a given
effective potential can be obtained by solving the Faddeev equations.
The full wave function can be decomposed into four components:
one for each two-body subcluster and one for the three-body 
cluster \cite{Meier:hj}.
For identical bosons, the three-body wave function is fully symmetric under 
exchange of particles and the Faddeev equations simplify considerably.
In this case, one only needs to solve equations involving  one of the 
two-body Faddeev components and the three-body component.
The two remaining two-body components can be obtained by permutations 
of particles. 

We follow Gl\"ockle and Meier \cite{Meier:hj} and decompose
the full three-body wave function as
\beq
\Psi=(1+P)\psi+\psi_{3}~,\qquad \mbox{where}\quad
 P=P_{13}P_{23}+P_{12}P_{23}
\label{eq:defP}
\eeq
is a permutation operator that generates the two not explicitly included
Faddeev components from $\psi$. The operator $P_{ij}$ simply
permutes particles $i$ and $j$. 
Since we are interested only in the binding energies and not in the
full wave function, we can eliminate the component $\psi_3$ and obtain
an equation for $ \psi$ alone:
\beq 
\psi=G_0\, t\, P\, \psi +G_0\, t\, G_0\, t_{3}\, (1+P)\,\psi~.
\label{faddeq}
\eeq
where $G_0$ denotes the free three-particle propagator.
Moreover, $t$ is the two-body t-matrix for the two-body subsystem 
described by the component $\psi$ and
$t_3$ is a auxiliary t-matrix 
defined by the solution of the three-body Lippmann-Schwinger equation with the
three-body effective interaction
\beq
V_{3}=|\xi\rangle\lambda_3\langle\xi|~,
\label{def:3pot}
\eeq
only. 
The three-body regulator function $|\xi\rangle$ is
defined as
\beq 
\langle {\bf u}_1, {\bf u}_2 |\xi\rangle=\exp\left(
-(u_1^2+3 u_2^2/4)/\Lambda^2\right)\,,
\eeq
where ${\bf u}_1$ and ${\bf u}_2$ are the usual Jacobi momenta.
Note that $t_3$ is only a technical construct that is
generally cutoff dependent and not observable. 
For the derivation of an explicit representation of Eq.~(\ref{faddeq})
in momentum space, we refer the reader to Refs.~\cite{Platter04,Lucthesis}.

The three-body binding energies are given by those values of $E$
for which Eq.~(\ref{faddeq}) has a nontrivial solution.
By expressing the two-body coupling constant $\lambda_2$ in terms
of the binding energy of the shallow two-body bound state, 
we have already renormalized the two-body problem.
The three-body problem can therefore be renormalized by requiring one of
the three-body binding energies to be fixed as one varies the cutoff.
All other low-energy three-body observables can then be predicted.
This renormalization procedure determines the three-body coupling constant
$\lambda_3(B_3,\Lambda)$ uniquely. 

The dimensionless coupling constant $\Lambda^4 \lambda_3(\Lambda)$
shows a limit cycle behavior \cite{Platter04}.
For large values
of the cutoff $\Lambda$, $\Lambda^4\lambda_3$ flows towards an 
ultraviolet limit cycle.
For $\Lambda\to \infty$, it has the limiting behavior
\beq
\lambda_3(\Lambda)=\frac{c}{\Lambda^4} \;\frac{\sin(s_0 \ln(\Lambda/L_3)
  -\arctan(1/s_0))}{\sin(s_0 \ln(\Lambda/L_3)+\arctan(1/s_0))}~,
\label{eq:limcyc}
\eeq
where $s_0\approx 1.00624$ is a transcendental number that determines the
period of the limit cycle. The constant $c=0.074 \pm 0.003$ was 
determined in Ref.~\cite{Platter04}. The discrete scaling symmetry
associated with a limit cycle is manifest in Eq.~(\ref{eq:limcyc}).
If the cutoff $\Lambda$ is multiplied by a factor $\exp(n\pi/s_0) 
\approx (22.7)^n$ with $n$ an integer, the 
three-body coupling $\lambda_3$ is unchanged. 
$L_3$ is a three-body parameter generated by dimensional transmutation. 
It can be determined by fixing
a three-body binding energy $B_3$.
Alternatively, one could use a three-body binding energy directly
to characterize the value of the three-body coupling $\lambda_3$ at a 
given cutoff \cite{Afnan:2003bs,Platter:2006ev}.

\subsection{The Four-Body Sector}
\label{sec:effT:4body}
We now turn to the four-body sector. The four-body binding energies
are given by the Yakubovsky equations 
which are a generalization of the Faddeev equations 
to the four-body system. The full four-body wave function $\Psi$ is 
first decomposed into Faddeev components, followed by a second decomposition 
into Yakubovsky components. For identical 
bosons, one has two Yakubovsky components $\psi_A$ and
$\psi_B$. We start from the Yakubovsky equations
including a general three-body force in the form written down by
Gl\"ockle and Kamada \cite{Glockle:1993vr}. The full four-body 
bound state wave function is decomposed into the Yakubovsky
components $\psi_A$ and $\psi_B$ via
\beq
\label{eq:yak_wavefunction}
\Psi=(1+(1+P) P_{34})(1+P)\psi_A +(1+P)(1+\tilde{P})\psi_B~,
\eeq
where $P_{ij}$ exchanges particles $i$ and $j$, $P$ is defined in 
Eq.~(\ref{eq:defP}), and $\tilde{P}$ is given by
\beq
\tilde{P}=P_{13}P_{24}~.
\eeq
The equations for the two wave function components read:
\bea
\psi_A &=& G_0 t_{12} P [ (1+P_{34})\psi_A+\psi_B]
+\frac{1}{3}(1+G_0 t_{12})G_0 V_{3} \Psi~, \nonumber \\
\psi_B &=& G_0 t_{12} \tilde{P} [ (1+P_{34})\psi_A+\psi_B]~,
\label{eq:yaku}
\eea
where $t_{12}$ denotes the two-body t-matrix for particles 1 and 2
and $V_3$ is the three-body force defined in Eq.~(\ref{def:3pot}).
Note that the three-body force couples to the 
full four-body wave function $\Psi$. The factor of one third in front 
of the three-body force term arises because we insert the full 
three-body interaction for $V_3$. This is possible
since we consider three-body contact interactions which are symmetric 
under the exchange of any pair of particles.

The renormalization analysis of the four-body system is complicated 
by the cutoff dependence of the number of  bound states in the three-body 
subsystems. The further the cutoff $\Lambda$ is increased, the more
three-body bound states appear. While this has no influence on low-energy 
three-body observables, it creates an instability in the four-body system
which can collapse into a deep three-body bound state plus another particle.
This limits cutoff variations to an interval $\Lambda_0 < \Lambda < 
22.7 \,\Lambda_0$ for some $\Lambda_0$, in which the number of three-body 
bound states remains constant. Despite this restriction, the cutoff can
be varied by more than a factor of ten which is
sufficient to study the renormalization properties and 
obtain converged numerical results \cite{Platter04}. 

In Ref.~\cite{Yama06}, it was pointed out that the calculation of
\cite{Platter04} for the $^4$He tetramer
was limited to repulsive three-body forces $\lambda_3 > 0$. 
The ability to renormalize without a four-body parameter was attributed
to a very small probability of the four particles to be close together
due to the repulsive three-body force. We note that 
the sign of the three-body force is cutoff dependent because
of the limit cycle behavior of $\lambda_3(\Lambda)$ in Eq.~(\ref{eq:limcyc}).
In Ref.~\cite{Platter04}, we have performed calculations with both
attractive and repulsive three-body forces.
In particular, the three-body force is attractive for cutoffs close 
to $\Lambda_0$.
However, in this case $\Lambda \sim (mB_4^{(0)})^{1/2}$ and 
the ground state energy $B_4^{(0)}$
is far from the converged value. The final converged value for $B_4^{(0)}$ 
was indeed obtained with a repulsive three-body force. 
The excited state energy $B_4^{(1)}$, however, is already converged
at smaller cutoffs where the three-body force is attractive.
Increasing the cutoff beyond $22.7 \,\Lambda_0$, where the three-body
force is attractive again would require to project out the unphysical 
deep three-body state that appears in this case.
Such a subtraction is beyond the scope of this work.
In the remainder of the paper, we will follow \cite{Platter04} and
not include a four-body parameter in our calculations.
\section{Four-Body Universality}
\label{sec:res}
We now apply this effective theory to calculate the universal properties of 
the 4-boson bound state spectrum and universal scaling functions.
We do not discuss the 2- and 3-boson spectrum in detail, since such
a discussion can be found in 
Refs.~\cite{Braaten:2004rn,Bedaque:1998km, Platter04}.
\subsection{Bound State Spectrum}
The four-body spectrum is most conveniently discussed in a
generalized Efimov plot. 
This plot was introduced by Vitaly Efimov to summarize the universal
properties of the three-body spectrum \cite{Efimov71,Braaten:2004rn}.
The set of all possible low-energy three-body states 
can be represented as points $(a^{-1}, K)$ on the plane 
whose horizontal axis is $1/a$ 
and whose vertical axis is the momentum variable
\begin{eqnarray}
K = {\rm sign} (E) \sqrt{m|E|} \,.
\label{K-def}
\end{eqnarray}
It is convenient to introduce polar coordinates consisting of
a radial variable $H$ and an angular variable $\xi$ defined by
\begin{eqnarray}
1/a &=& H \cos \xi \,,
\qquad
K   = H \sin \xi \,.
\label{Hxi-def}
\end{eqnarray}
In terms of these polar coordinates,
the discrete scaling symmetry in the three-body system
is simply a rescaling  of the radial variable: $H \to \exp(n\pi/s_0)\; H$.

\begin{figure}[htb]
\bigskip
\centerline{\includegraphics*[width=8cm,angle=0]{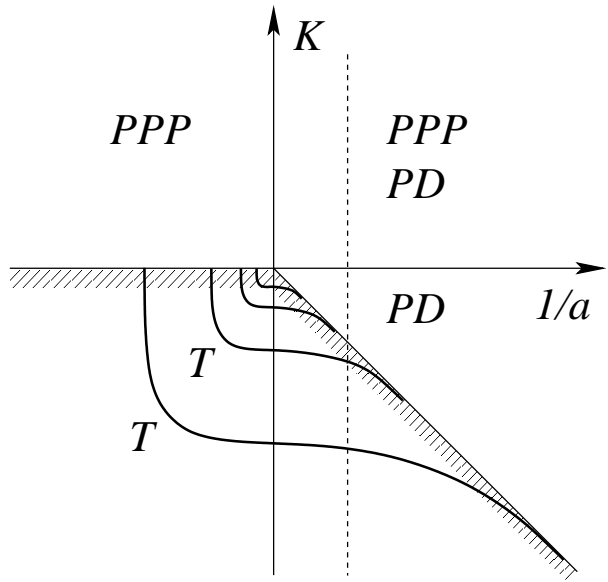}}
\medskip
\caption
{The $a^{-1}$--$K$ plane for the three-body problem. The allowed regions for
three-particle scattering states and particle-dimer scattering states are
labeled $PPP$ and $PD$, respectively. The heavy lines labeled $T$
are two of the infinitely many branches of Efimov states. 
The cross-hatching indicates the threshold for scattering states. 
}
\label{fig:3body}
\end{figure}

The $a^{-1}$--$K$ plane for three identical bosons 
is shown in Fig.~\ref{fig:3body}. 
We will refer to the bosons simply as particles in the following.
The possible states are three-particle scattering states ($PPP$),
particle-dimer scattering states ($PD$), and Efimov trimers ($T$).
The threshold for scattering states is indicated by the hatched area.
The Efimov trimers are represented by the heavy
lines below the threshold.\footnote{The curves for the trimer 
binding energies in Fig.~\ref{fig:3body}
actually correspond to plotting $H^{1/4}\sin\xi$ versus $H^{1/4}\cos\xi$.
This effectively reduces the discrete symmetry
factor 22.7 down to $22.7^{1/4} = 2.2$, allowing a greater range of
$a^{-1}$ and $K$ to be shown in the Figure.} 
There are infinitely many branches of Efimov trimers close to threshold, 
but only a few are shown.
A given physical system has a specific value of the scattering length
and can be represented by a vertical line.
The infinite scattering length limit
corresponds to tuning this line to the $K$ axis.
For a real physical system, not all states will behave as depicted in the 
figure.  With momenta of order $1/l$ (or energies of order
$1/(ml^2)$) one is able to probe details
of the short-distance mechanism leading to the large scattering
length and the effective theory does no longer apply.
As a consequence, only states with $Kl \ll 1$ and 
systems with $l/|a| \ll 1$ will show the universal behavior.

\begin{figure}[tb]
\centerline{\includegraphics*[width=5.in,angle=0]{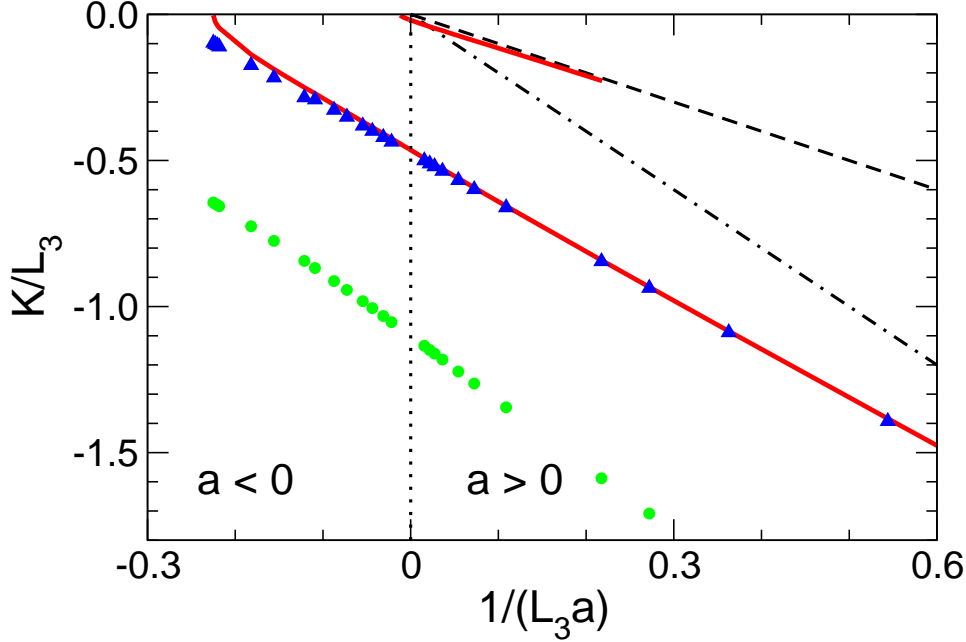}}
\caption{\label{fig:efiplot} The $a^{-1}$--$K$ plane for the four-body 
problem. The circles and triangles indicate the 
four-body ground and excited state energies $B_4^{(0)}$ and $B_4^{(1)}$,
while the lower (upper) solid lines give the thresholds for decay into a 
ground state (excited state) trimer and
a particle. The dash-dotted (dashed) lines give the thresholds for
decay into two dimers (a dimer and two particles).
The vertical dotted line indicates infinite scattering length.
All quantities are given in units of the three-body parameter $L_3$.
}
\end{figure}

In Fig.~\ref{fig:efiplot}, we generalize this plot to the four-body system
and plot the square root of the four-body energy versus the inverse
scattering length.
The lines representing the trimer energies in Fig.~\ref{fig:3body}
now become the scattering thresholds for trimer-particle scattering. 
The vertical dotted line indicates the limit of infinite scattering length.
Our results for the
four-body ground and excited state energies $B_4^{(0)}$ and $B_4^{(1)}$
are given by the circles and triangles, respectively.
Four-body states (tetramers) can only be stable if their energy is below 
all scattering thresholds, otherwise they become resonances and 
aquire a width from the decay into the corresponding scattering states.
The lower (upper) solid lines indicate the thresholds for scattering of a
ground state (excited state) trimer and another particle. For positive $a$,
there are also scattering thresholds for scattering of two dimers and 
a dimer plus two particles indicated by the dash-dotted and dashed
lines, respectively. These thresholds apply to both tetramer states.
A given physical system is
again characterized by a vertical line corresponding to a particular
value of the scattering length. Depending on the value of the 
scattering length, the ground state trimer-particle threshold or 
the dimer-dimer
threshold can be closest to the tetramer states. In this paper, we
focus on the region where the closest threshold is given by ground state
trimer-particle scattering. If the plot was extended further to the right,
the dimer-dimer threshold will eventually become the lowest one.

We have obtained the dependence of the four-body energies on the 
scattering length $a$ for fixed three-body parameter $L_3$ from explicit  
solutions of the Yakubovsky equations (\ref{eq:yaku}).
All quantities are given  in units of $L_3$. Note that $L_3$ is
only defined up to a rescaling by the discrete scaling factor
$\exp(n\pi/s_0)\approx (22.7)^n$, so the absolute scale of 
Fig.~\ref{fig:efiplot} is arbitrary up to such factors. 
The circles and triangles show our results for the 
four-body ground and excited state energies $B_4^{(0)}$ and $B_4^{(1)}$
as a function of the scattering length, respectively.
The tetramer excited state remains close to the threshold
for decay into a trimer and another particle for essentially all
calculated values of $1/a$. It moves away from this threshold 
near the point  where the threshold disappears at $K=0$.
The tetramer ground state is considerably deeper. It stays essentially
parallel to the excited state but starts to move away for small positive
scattering lengths. 
Due to numerical instabilities, we were not able to reach
the region where the two-dimer threshold becomes relevant.

For the limits of universal behavior the same restrictions as
in the three-body system apply.
With momenta of order $1/l$ (or energies of order
$1/(ml^2)$) one is able to probe details
of the short-distance mechanism leading to the large scattering
length and the effective theory does no longer apply.
As a consequence, only states with $Kl \ll 1$ and 
systems with $l/|a| \ll 1$ will show the universal behavior 
in Fig.~\ref{fig:efiplot}. Which states are within the universal window
then depends on the actual values of $a$ and $l$ in the system under 
consideration. For $^4$He atoms
there is some evidence that both tetramer states are within this 
range \cite{Platter04}.

It is interesting to note that the universal theory
always produces two  four-body states \cite{Lucthesis}. 
The number of four-body states is independent of the number of three-body
states. The latter can be expected since the excited three-body states are 
shallower by at least a factor of 515 and therefore should have 
little influence on the four-body states. The discrete scaling
symmetry then suggests that there are exactly two four-body resonances
between any two three-body states with the shallower of the four-body states
having a slightly lower energy than the shallower of the two three-body 
states. These states are stable for some range of cutoffs, but become 
resonances with the appearance of a deeper three-body state when the cutoff
is increased beyond the critical value $22.7 \Lambda_0$ discussed
in section  \ref{sec:effT:4body}.
Checking this conjecture would require finding resonance poles
in the complex plane which is beyond the scope of this work.

\subsection{Universal Correlations}
We now consider the universal correlations in the three-body system for
negative scattering length. In particular, we focus on the correlation
between the binding energies in the three- and four-body systems.
Such correlations were first observed
in nuclear physics and are known as Tjon lines \cite{Tjo75}.
For recent discussions of the nuclear Tjon line in the context
of the  EFT for short-range interactions
and low-momentum nucleon-nucleon potentials, 
see Refs.~\cite{Platter:2004zs,Nogga:2004ab}.
The Tjon lines for spinless bosons with
positive scattering length were discussed in 
\cite{Platter04} for the range of binding energies relevant to 
$^4$He atoms.

\begin{figure}[tb]
\centerline{\includegraphics*[width=5.in,angle=0]{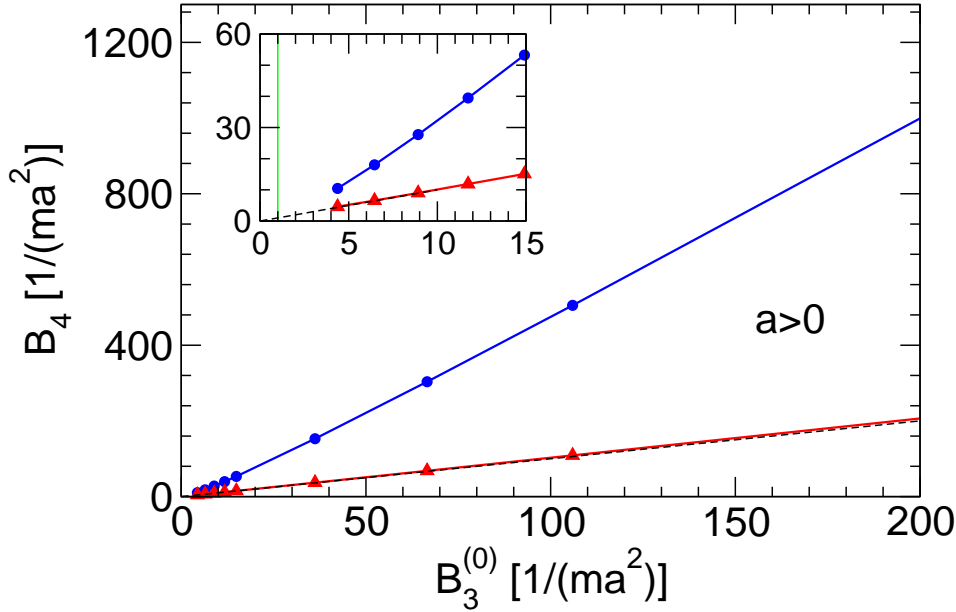}}
\caption{\label{fig:tjon-apos} The correlation between the 
three-body ground state energy $B_3^{(0)}$ and the four-body
ground and excited state energies $B_4^{(0)}$ (circles)
and $B_4^{(1)}$ (triangles)
in units of $1/(ma^2)$ for $a>0$. The inset shows the threshold
region in more detail.}
\end{figure}

\begin{figure}[tb]
\centerline{\includegraphics*[width=5.in,angle=0]{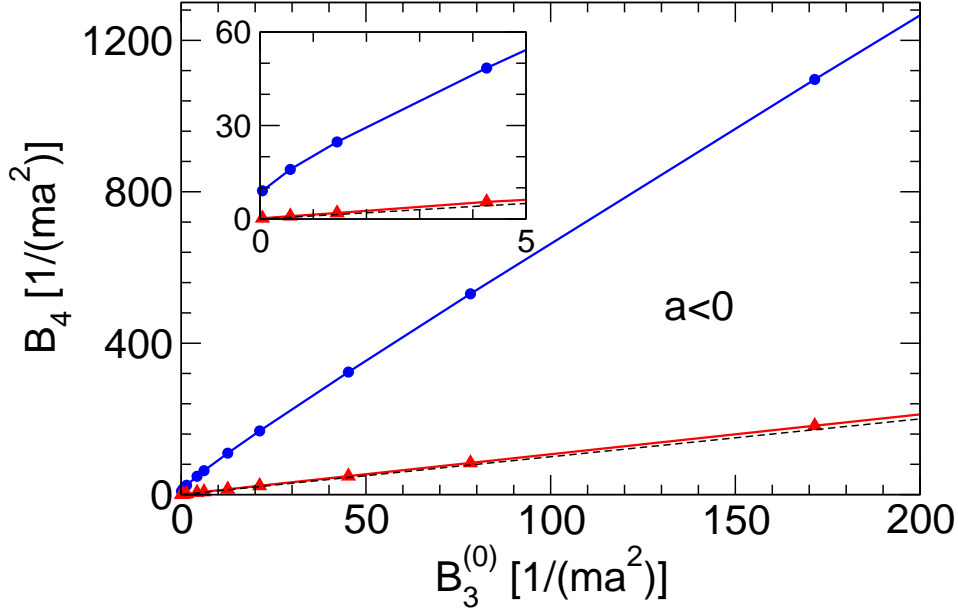}}
\caption{\label{fig:tjon-aneg} The correlation between the 
three-body ground state energy $B_3^{(0)}$ and the four-body
ground and excited state energies $B_4^{(0)}$ (circles) and 
$B_4^{(1)}$ (triangles)
in units of $1/(ma^2)$ for $a<0$. The inset shows the threshold
region in more detail.
}
\end{figure}

Here, we discuss the correlation between the three-body ground state
energy $B_3^{(0)}$ and the four-body
ground and excited state energies $B_4^{(0)}$ and $B_4^{(1)}$ for values
of $B_3^{(0)}$ from threshold up to $B_3^{(0)}\approx 200/(ma^2)$.
In Fig.~\ref{fig:tjon-apos}, we show the correlation between the 
ground state energy in the three-body system and the 
ground and excited state energies in the four-body system for $a>0$. 
The corresponding plot for negative scattering length is shown in 
Fig.~\ref{fig:tjon-aneg}.

The inset in both figures shows the threshold region in more detail. 
The nonlinear behavior of the Tjon lines in this region is evident.
For positive scattering length, we can not calculate
all the way down to the three-body threshold at $B_3=1/(ma^2)$
due to numerical instabilities.
The correlations have positive curvature for $a>0$ and negative
curvature for $a<0$.
Outside the threshold region, the correlation is approximately linear
independent of the sign of $a$.
A similar linear relation holds for the correlation 
between different three-body energies (e.g. ground and excited states)
\cite{Braaten:2004rn}. This is a direct consequence of the discrete
scaling symmetry in the three-body system. For $1/a=0$, this 
symmetry ensures that the binding energies of subsequent three-body
states differ by factors of approximately 515. If the scattering length
is finite, the correlation is still linear to a good approximation.
Only for states close to threshold the linearity breaks down. In this case,
there is still an exact scaling symmetry, but it relates states
corresponding to different values of the scattering length.
Since no new parameter enters at leading order in the four-body system,
the above arguments immediately generalize to the four-body case.
Away from the thresholds, the binding energies of the ground
and excited four-body states are related to the deepest three-body
state by factors of 5 and 1.01, respectively. 

Similar linear correlations between the binding energies of
$N$- and $(N-1)$-body systems with positive scattering length
and $N>4$ were recently found 
by Hannah and Blume \cite{Hanna06}. They studied bosonic clusters
with up to $N=40$ atoms interacting additively through two-body
van der Waals potentials. Using exact Monte Carlo methods, they
determined the energies as well as the interparticle distances.
They found approximately linear relations for the correlation
between the ground state energy ratios $B^{(0)}_{N}/B^{(0)}_{N-2}$ and 
$B^{(0)}_{N-1}/B^{(0)}_{N-2}$ for $N$ up to $10$. This observation
suggests that the discrete scaling symmetry holds up to $N=10$. 
Consequently, no new parameters would enter into the effective theory 
at leading order for up to 10 particles.
If true, it would open up the exciting possibility to predict 
the universal properties of $N$-body system up to $N=10$ from
the scattering length $a$ and the three-body parameter $L_3$ alone.

\section{Summary and Outlook}
\label{sec:conc}
In this paper, we have studied the universal
properties of the four-body system with large scattering length. 
These properties are useful for understanding the cluster structure 
of nuclei \cite{Mar06} as part of the planned experimental program at FAIR
and experiments on the creation of weakly-bound tetramers with cold atoms
close to a Feshbach resonance.
We have concentrated on the bound state problem of four bosons 
starting from the Yakubovsky equations.
We have constructed an effective interaction potential including 
both a two- and three-body contact interaction.
This is the minimal set of contact interactions required for 
renormalization of the three-body problem.
The two parameters of the effective potential were determined from
matching to the binding energy of the dimer ($a>0$) or 
the two-body scattering length ($a<0$) and the excited state
of the trimer. We have then solved the four-body bound state problem
and obtained the scattering length dependence of the four-body 
bound state spectrum. This dependence was presented in a generalized
Efimov plot. For all considered values of $a$, 
we have found two four-body states. 
The tetramer excited state stays close to the trimer-particle threshold,
while the ground state is considerably deeper.
We have conjectured, that there are always two four-body resonances
between any two three-body states. As the cutoff is increased, these
states come to life as stable bound states, but turn into resonances
as additional deep three-body states appear when the cutoff is increased
beyond a critical value.

An important consequence of the large scattering length is the 
emergence of universal scaling functions. Since there are only two 
parameters at leading order, $a$ and $L_3$, few-body observables
normalized to the scattering length $a$ must be correlated and follow
a line parameterized by $L_3$.
We have calculated the universal scaling functions relating 
the tetramer energies to the trimer ground state energy. A
similar correlation between the triton and alpha particle 
energies is known from nuclear physics as the Tjon line 
\cite{Tjo75,Platter:2004zs,Nogga:2004ab}.
Close to the three-body thresholds, these correlations have 
positive curvature for $a>0$ and negative curvature for $a<0$.
Away from the thresholds, they are linear
to very high accuracy. This linearity can be understood from the
discrete scaling symmetry. Similar linear correlations between the 
binding energies of $N$- and $(N-1)$-body systems with $N>4$ were 
recently found from exact Monte Carlo calculations using 
two-body van der Waals potentials \cite{Hanna06}. 
This observation suggests the absence of $N$-body parameters for $N>4$,
which would allow to predict $N$-body observables from two- and
three-body input alone.

The limits of universal behavior in real systems are determined by
the values of $a$ and $l$ in those systems. 
With momenta of order $1/l$ (or energies of order
$1/(ml^2)$) one is able to probe details
of the short-distance mechanism leading to the large scattering
length and our effective theory does no longer apply.
As a consequence, only states with $Kl \ll 1$ and 
systems with $l/|a| \ll 1$ will show the universal behavior 
discussed in this paper. Corrections are suppressed by powers 
of $l/|a|$ and $\kappa l$ where $\kappa$ is the typical momentum
scale in the process considered (e.g.~the binding momentum for bound 
states). In Ref.~\cite{Platter04}, we have shown that our results
can be applied to the $^4$He tetramer. We are not aware of any 
data on four-body systems with large negative scattering length.

We also stress that the universality of our results relies on the 
validity of the renormalization group analysis in Ref.~\cite{Platter04}.
In this work, renormalization of the four-body equations was
achieved without the introduction of a four-body interaction.
If this result would not
hold, the properties of the four-body system with large scattering
length would also depend on a four-body parameter
(cf.~Ref.~\cite{Yama06}).

There are a number of directions that should be pursued in future work.
First, it would be interesting to extend the calculation to 
smaller positive values of $a$, where the relevant decay threshold for 
the tetramer states is the dimer-dimer threshold. This would
complete the generalized Efimov plot. Our conjecture of two
four-body resonances between any two trimer states could be tested
by extending the calculation to complex values of the energy. The 
widths of such states are currently unknown.
Moreover, the application of our results to cluster structures
in nuclei requires the inclusion of the long-range Coulomb force.
Work in this direction is in progress.
The general power counting for four-body forces is still not understood.
At which order does the leading four-body interaction enter? In the 
three-body system, e.g., the first and second order correction are 
due to the two-body effective range \cite{Platter:2006ev}. 
If a similar situation holds in the four-body system, 
it would be possible to predict low-energy four-body observables up to 
corrections of order $(l/|a|)^3$ from two- and three-body information alone. 
The extension of the effective theory to calculate four-body scattering
observables would be very valuable. 
The knowledge of the dimer-dimer scattering length, for example,
is important for experiments with ultracold atoms. For the simpler
problem of fermions with two spin states (where the three-body parameter
$L_3$ does not contribute), the dimer-dimer scattering length
was recently calculated exactly \cite{Petrov04}
and using a perturbative $\epsilon$-expansion around $d=4$ \cite{Rupak06}.

\begin{acknowledgments}
We thank Eric Braaten for discussions.
This research was supported by the U.S. Department of Energy under 
grant DE-FG02-93ER40756, by the DFG through funds provided
to the SFB/\-TR 16 \lq\lq Subnuclear structure of matter'', and by the
BMBF under contract number 06BN411.
\end{acknowledgments}


\begin{thebibliography}{99}

\bibitem{Bedaque:2002mn}
P.~F.~Bedaque and U.~van Kolck,
Ann.\ Rev.\ Nucl.\ Part.\ Sci.\  {\bf 52}, 339 (2002)
[arXiv:nucl-th/0203055].

\bibitem{Epelbaum:2005pn}
  E.~Epelbaum,
  Prog.\ Part.\ Nucl.\ Phys.\  {\bf 57}, 654 (2006)
  [arXiv:nucl-th/0509032].

\bibitem{Braaten:2004rn}
  E.~Braaten and H.-W.~Hammer,
  Phys.\ Rept.\  {\bf 428}, 259 (2006)
  [arXiv:cond-mat/0410417].

\bibitem{Efimov71}
V.~Efimov,
        Yad.\ Fiz.\ {\bf 12}, 1080 (1970)
        [Sov.\ J.\ Nucl.\ Phys. {\bf 12}, 589 (1971)].

\bibitem{Efimov79}
V.~Efimov,
        Yad.\ Fiz.\ {\bf 29}, 1058 (1979)
        [Sov.\ J.\ Nucl.\ Phys.\ {\bf 29}, 546 (1979)].


\bibitem{Bedaque:1998kg}
P.F.~Bedaque, H.-W.~Hammer and U.~van Kolck,
Phys.\ Rev.\ Lett.\  {\bf 82}, 463 (1999)
[arXiv:nucl-th/9809025].

\bibitem{Bedaque:1998km}
P.F.~Bedaque, H.-W.~Hammer and U.~van Kolck,
Nucl.\ Phys.\ A {\bf 646}, 444 (1999)
[arXiv:nucl-th/9811046].

\bibitem{Mohr:2005pv}
  R.F.~Mohr, R.J.~Furnstahl, R.J.~Perry, K.G.~Wilson and H.-W.~Hammer,
  Ann.\ Phys.\  {\bf 321}, 225 (2006)
  [arXiv:nucl-th/0509076].

\bibitem{Afnan:2003bs}
I.~R.~Afnan and D.~R.~Phillips,
Phys.\ Rev.\ C {\bf 69}, 034010 (2004)
[arXiv:nucl-th/0312021].

\bibitem{Platter:2006ev}
  L.~Platter and D.R.~Phillips,
  Few Body Syst.\  {\bf 40}, 35 (2006)
  [arXiv:cond-mat/0604255].

\bibitem{Platter:2006ad}
  L.~Platter,
  Phys.\ Rev.\ C {\bf 74}, 037001 (2006)
  [arXiv:nucl-th/0606006].

\bibitem{Platter04}
L.~Platter, H.-W.~Hammer, and Ulf-G.~Mei\ss ner,
Phys.\ Rev.\ A {\bf 70}, 052101 (2004) [arXiv:cond-mat/0404313].

\bibitem{Lepage:1997cs}
G.~P.~Lepage,
arXiv:nucl-th/9706029.

\bibitem{Mohr:2003du}
R.F.J.~Mohr,
arXiv:nucl-th/0306086.

\bibitem{Platter:2004zs}
  L.~Platter, H.-W.~Hammer and U.-G.~Mei\ss ner,
  Phys.\ Lett.\ B {\bf 607}, 254 (2005)
  [arXiv:nucl-th/0409040].

\bibitem{Nogga:2004ab}
  A.~Nogga, S.~K.~Bogner and A.~Schwenk,
  Phys.\ Rev.\  C {\bf 70}, 061002 (2004)
  [arXiv:nucl-th/0405016].

\bibitem{Gibs76}
B.F.~Gibson and D.R.~Lehman, Phys.\ Rev.\ C {\bf 14}, 685 (1976).

\bibitem{Fons76}
A.C.~Fonseca and P.E.~Shanley, Phys.\ Rev.\ C {\bf 14}, 1343 (1976).

\bibitem{Tjon78}J.A.~Tjon, in \lq\lq Few-body systems and nuclear forces II'',
edited by H.~Ziegel et al. (Springer, Berlin, 1978).

\bibitem{Lim84}T.K.~Lim, Nucl.\ Phys.\ A {\bf 416}, 491 (1984).

\bibitem{Kama01}H.~Kamada et al.,  Phys.\ Rev.\ C {\bf 64}, 044001 (2001).

\bibitem{Blume:2000}
D.~Blume and C.H.~Greene, J.\ Chem.\ Phys.\ {\bf 112}, 8053 (2000).

\bibitem{Yama06}
M.T.~Yamashita, L.~Tomio, A.~Delfino, T.~Frederico,
Europhys.\ Lett.\ {\bf 75}, 555 (2006)
[arXiv:cond-mat/0602549].

\bibitem{Mar06}
J.A.~Maruhn, M.~Kimura, S.~Schramm, P.-G.~Reinhard, H.~Horiuchi,
and A.~Tohsaki, 
Phys.\ Rev.\ C {\bf 74}, 044311 (2006)
[arXiv:nucl-th/0606053].

\bibitem{Chin05}
C.~Chin, T.~Kraemer, M.~Mark, J.~Herbig, P.~Waldburger, 
H.-C.~N\"agerl, and R. Grimm, 
Phys.\ Rev.\ Lett.\ {\bf 94}, 123201 (2005).

\bibitem{Kaplan:1998tg}
D.B.~Kaplan, M.J.~Savage and M.B.~Wise,
Phys.\ Lett.\ B {\bf 424}, 390 (1998)
[arXiv:nucl-th/9801034].

\bibitem{vanKolck:1998bw}
U.~van Kolck,
Nucl.\ Phys.\ A {\bf 645}, 273 (1999)
[arXiv:nucl-th/9808007].

\bibitem{Hammer:2001gh}
H.-W.~Hammer and T.~Mehen,
Phys.\ Lett.\ B {\bf 516}, 353 (2001)
[arXiv:nucl-th/0105072].

\bibitem{Bedaque:2002yg}
P.F.~Bedaque, G.~Rupak, H.W.~Grie\ss hammer and H.-W.~Hammer,
Nucl.\ Phys.\ A {\bf 714}, 589 (2003)
[arXiv:nucl-th/0207034].

\bibitem{Griesshammer:2004pe}
  H.W.~Grie\ss hammer,
  Nucl.\ Phys.\ A {\bf 744}, 192 (2004)
  [arXiv:nucl-th/0404073].

\bibitem{Lucthesis}
L. Platter, Ph.D. Thesis, Bonn University (2005).

\bibitem{Ziegelmann}
E.W.~Schmid and H.~Ziegelmann, {\it The Quantum Mechanical Three-Body
  Problem}, (Vieweg, 1971).

\bibitem{Meier:hj}
W.~Meier and W.~Gl\"ockle,
Phys.\ Rev.\ C {\bf 28}, 1807 (1983).

\bibitem{Glockle:1993vr}
W.~Gl\"ockle and H.~Kamada,
Nucl.\ Phys.\ A {\bf 560}, 541 (1993).

\bibitem{Tjo75}J.A.~Tjon, 
Phys.\ Lett.\ B {\bf 56}, 217 (1975).

\bibitem{Hanna06}
G.J.~Hanna and D.~Blume, 
Phys.\ Rev.\ A {\bf 74}, 063604 (2006)
[arXiv:cond-mat/0607519].


\bibitem{Petrov04} D.S.\ Petrov, C.\ Salomon, and G.V.\ Shlyapnikov,
Phys.\ Rev.\ Lett.\ {\bf 93}, 090404 (2004)
[arXiv:cond-mat/0309010].

\bibitem{Rupak06}
G.~Rupak, arXiv:nucl-th/0605074.

\end{thebibliography}
\end{document}